\documentclass[times]{aastex631}
\newcommand{\nc}{\newcommand}
\newcommand{\HII}{H {\sc ii}}
\nc{\msun}{\ensuremath{\mathrm{M}_\odot}}
\nc{\lsun}{\ensuremath{\mathrm{L}_\odot}}
\nc{\thCO}{$^{13}$CO}
\nc{\CeiO}{C$^{18}$O}
\nc{\kms}{\mbox{km~s$^{-1}$}}
\nc{\Kkms}{\mbox{K\,km~s$^{-1}$}}

\nc{\twCO}{$^{12}$CO}
\def\O0{C$^{0}$}

\nc{\cmsq}{\mbox{cm$^{-2}$}}
\nc{\cmcub}{\mbox{cm$^{-3}$}}
\nc{\tkin}{\mbox{$T_{\rm kin}$}}

\newcommand{\CII}{[C {\sc ii}]}
\newcommand{\OI}{[O {\sc i}]}
\def\ptsec{$''\mskip-7.6mu.\,$}
\def\cplus{C$^{+}$}
\nc{\thCII}{[$^{13}$C {\sc ii}]}
\def\vlsr{v$_{\rm LSR}$}

\submitjournal{APJ}

\shorttitle{Kinematics of PDR in M20}
\shortauthors{B. Mookerjea \& G. Sandell}
\usepackage{graphicx}	% Including figure files
\usepackage{amsmath}	% Advanced maths commands
\usepackage{url}
\usepackage{hyperref}
\hypersetup{
    colorlinks=true,
    citecolor=blue,
    linkcolor=blue,
    filecolor=blue,      
    urlcolor=blue,
    pdftitle={Overleaf Example},
    pdfpagemode=FullScreen,
    }
\usepackage{comment}

\begin{document}
\graphicspath{{./}{figures/}}
%% This is the end of the preamble.  Indicate the beginning of the
%% manuscript itself with \begin{document}.

%%%%%%%%%%%%%%%%%%%%%%%%%%%%%%%%%%%%%%%%%%%%%%%%%%

%%%%%%%%%%%%%%%%%%% TITLE PAGE %%%%%%%%%%%%%%%%%%%

% Title of the paper, and the short title which is used in the headers.
% Keep the title short and informative.

\title[Kinematics of PDR in M20]{Tracing the layers of Photodissociated gas in Trifid Nebula}

% The list of authors, and the short list which is used in the headers.
% If you need two or more lines of authors, add an extra line using \newauthor
%\correspondingauthor{Bhaswati Mookerjea}
%\email{bhaswati@tifr.res.in}

\author[0000-0003-1766-6303]{Bhaswati Mookerjea}
\affiliation{Department of Astronomy \& Astrophysics, Tata Institute of
Fundamental Research,\\ Homi Bhabha Road, Mumbai 400005, India}{\email{bhaswati@tifr.res.in}}
\author[0000-0003-0121-8194]{G\"oran Sandell}
\affiliation{Institute for Astronomy, University of Hawai`i at Manoa, Hilo, \\ 640 N. Aohoku Place, Hilo, HI 96720, USA}

%%%%% AUTHORS - PLACE YOUR OWN COMMANDS HERE %%%%%

% Please keep new commands to a minimum, and use \newcommand not \def to avoid
% overwriting existing commands. Example:
%\newcommand{\pcm}{\,cm$^{-2}$}	% per cm-squared

\begin{abstract}
Photodissociated gas bears the signature of the dynamical evolution of the
ambient interstellar medium impacted by the mechanical and radiative feedback
from an expanding \HII\ region. Here we present an analysis of the kinematics of
the young Trifid nebula, based on velocity-resolved observations of the far-infrared fine-structure lines of \CII\ at 158\,\micron\ and \OI\ at  63\,\micron. The distribution of the photodissociated regions (PDRs) surrounding the nebula is consistent with a shell-like structure created by the \HII\ region expanding with a velocity of 5\,\kms. Comparison of ratios of \CII\ and \OI 63\,\micron\ intensities for identical velocity components with PDR models indicate a density of 10$^4$\,\cmcub. The red- and blue-shifted PDR shells with a combined mass of 516\,\msun\ have a kinetic energy of $\sim 10^{47}$\,erg. This is consistent with the thermal energy of the \HII\ region as well as with the energy deposited by the stellar wind luminosity from HD 169442A, an O7 V star, over the 0.5\,Myr lifetime of the star.  The observed momentum of the PDR shell is lower than what theoretical calculations predict for the radial momentum due to the shell being swept up by an expanding \HII\ region, which suggests that significant mass loss has occurred in M20 due to the dispersal of the surrounding gas by the advancing ionization front.
\end{abstract}


%%%%%%%%%%%%%%%%% BODY OF PAPER %%%%%%%%%%%%%%%%%%
\section{Introduction} 

\begin{figure} \centering
\includegraphics[width=9.cm]{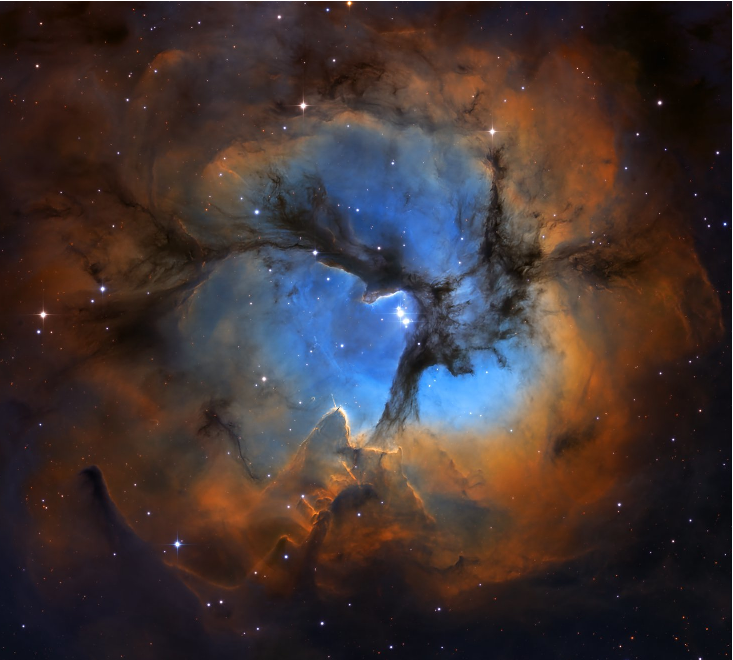} \caption{Composite three-color optical image of M20 observed by Wolfgang Promper from his remote observatory in Namibia with narrowband filters, OIII ($blue$), H$\alpha$ ($green$) and SII ($red$). The authors have obtained the permission to use the image available at \url{https://www.astrobin.com/2c61wh/B/?q=\%22HD164294\%22} 
\label{fig_m20op}} \end{figure}

The Trifid nebula (M20) is an \HII\ region with an age of approximately 0.5\,
Myr \citep{Lefloch2000} ionized by the O7 V star HD164492A \citep{Lynds1985} 
located in the Sagittarius spiral arm at a distance of 1180$\pm$25pc
\citep{Kuhn2021}.  Figure\,\ref{fig_m20op} shows a composite three-color optical image of M20 observed by Wolfgang Promper from his remote observatory in Namibia with narrowband filters, O III ($blue$), H$\alpha$ ($green$) and S II ($red$).  The image shows both the large-scale filamentary dust feature and the denser gas towards
the south-west that lie in front of the ionized gas in the \HII\ region
created by HD\,164492A.

The Infrared Space Observatory (ISO) and Hubble Space Telescope (HST)
observations \citep{Cernicharo1998,Lefloch2000} show the Trifid to be a dynamic
star-forming region containing young stellar objects (YSOs), many of them
undergoing episodes of violent mass ejection or transferring mass and energy to
the nebula in the form of jets. Comparison of a high-resolution radio continuum
image at 20\,cm and a CO map suggest that in its expansion the \HII\ region
has encountered a large dense molecular cloud to the south-west (SW), possibly
triggering the formation of new stars \citep{Cernicharo1998}.  From the Spitzer
Infrared Array Camera (Spitzer/IRAC) and Multi-Band Imaging Photometer for
Spitzer (MIPS) images of the Trifid nebula (M20), \citet{Rho2006} identified
about 160 YSOs at different evolutionary stages distributed around the whole
nebula, and beyond. These cover all stages of proto-stellar evolution, from 
Class 0 protostars to more evolved young stellar objects (YSOs). The youngest (Class 0/I protostars TC0--TC8) of the sources in the region are located within the dense far-IR and millimetre cores identified in the 1.3\,mm continuum map observed by
\citet{Lefloch2002}. In some of the dense cores evidence of jets 
originating from protostars have been discovered through narrow-band line
imaging in the optical with the HST. Sticking out of the
boundaries of one of the cores, TC2 is a 22-arcsec-long irradiated jet HH 399,
ending close to a bow shock $\sim 1.5$\arcmin\  away, in a low density environment
\citep{Cernicharo1998,Rosado1999, Yusef2005a}. Millimeter-wavelength molecular-line observations toward the nebula reveal
several cloud components at velocities of v$_{\rm LSR} \sim$ 2\,\kms\
(corresponding to the dark lanes crossing the optical nebula), v$_{\rm LSR}\sim
8$\,\kms\ (corresponding to dark features at the edge of the optical nebula),
and v$_{\rm LSR}\sim 18$\,\kms\
\citep{Kuhn2022,Cernicharo1998,Torii2011,Torii2017}. A bridging feature between
the 2 and 8\,\kms\ clouds is seen coincident with the optical nebula and indicates that these clouds are physically interacting
and may have collided \citep{Haworth2015,Torii2017,Kalari2021}. Alternatively,
the cloud kinematics could be interpreted as either an expanding \HII\ region or
turbulence \citep{Fukuda2000,Rho2008}. \citet{Kuhn2022} used the  \thCO(2--1)
channel maps to suggest that the 18\,\kms\ cloud which is of much larger spatial
extent is likely to be in the background. Until now studies of photodissociated gas in M20 have only been for a small region around the TC2 globule using  the far- and mid-infrared fine-structure lines detected with the LWS and SWS onboard ISO \citep{Lefloch2002}.

Here we present a study of the photodissociated gas in an approximately
400\arcmin$\times$400\arcmin\ region using the far-infrared (FIR)
lines of \CII\  at 158 $\mu$m and \OI\  at 63\,\micron\ that account for most of the cooling. The datasets used here not only cover a much larger area compared to that studied by \citet{Lefloch2002} but also have the advantage of having higher spatial and
more importantly spectral resolution compared to the ISO data. Recent
velocity-resolved studies of Galactic PDRs have shown that both the \CII\ and
\OI 63\,\micron\ emission are often moderately optically thick and sometimes
significantly self-absorbed \citep{Graf2012, Mookerjea2018, Guevara2020, Mookerjea2021}. We use the velocity-resolved \OI 63\,\micron\ and \CII\ spectra to
characterize the properties and distribution of the PDR gas vis-a-vis location
of these components with respect to the ionizing source by  segregating the
major velocity components of both the diffuse and the dense PDR gas.

\section{Observational Datasets}

For this work, we have used several sets of publicly available data, which are described below.

\subsection{GREAT/SOFIA observations}

We have retrieved observations  of the $^2$P$_{3/2}$$\rightarrow$$^2$P$_{1/2}$
fine structure transition of ionized carbon (\cplus) at 1900.5369\,GHz
(157.74\,\micron)  and the $^3$P$_{1}\rightarrow ^3$P$_{0}$ transition of atomic
oxygen, \OI 63\,\micron, at 4,744.77749\,THz of a region centered at the
location of the TC2 globule (($\alpha_{J2000}$: $18^h~02^m~27.22^s$,
$\delta_{J2000}$: $-23^\circ 04^\prime$ 27\ptsec5) in M20 from the data archive
of the Stratospheric Observatory for Infrared Astronomy
\citep[SOFIA;][]{young2012}. The observations (Id: 06\_0041; PI: M.  Kaufman)
were carried out using the German REceiver for Astronomy at Terahertz
frequencies \citep[GREAT;][]{heyminck2012} on 2018 June 08.  The GREAT was in
the Low Frequency Array/High Frequency Array (LFA/HFA) configuration with both
arrays operating in parallel. The LFA was tuned to \CII\ and the HFA was tuned
to \OI 63\,\micron. The beam sizes for the \CII\ and \OI 63\,\micron\ were 14\farcs1
and 6\farcs3 respectively.  The \CII\ map extends over a region of 
400\arcsec$\times$400\arcsec, while the usable part of the \OI 63\,\micron\ is
closer to 300\arcsec $\times$ 300\arcsec. We used the Level 3b data available on
the SOFIA archive for the region. All data were resampled to a spectral
resolution of 0.5\,\kms\ resulting in a median rms of 0.5\,K for both \CII\
and \OI 63\,\micron\ data.  The \OI 63\,\micron\ map was smoothed with a Gaussian kernel and regridded to the same spatial resolution as the \CII\ map, i.e. 14\farcm1(Fig.\,\ref{fig_intmaps}).

\subsection{\twCO(3--2) and HCO$^+$(4--3) \& (1--0)}

For comparison with our observations, we used maps of the $J$=3--2 transition of
 \twCO\ at 345795.9899\,MHz at and the $J$=4--3 transition of HCO$^+$ at 356734.2230\,MHz, both observed with the James Clerk Maxwell Telescope (JCMT) using the Heterodyne Array Receiver Program (HARP) receiver with a beam size of 14\arcsec. The observations were performed as part of the proposal M07AH24A. The spectra were downloaded directly from the JCMT archive at the Canadian Astronomical Data Centre (CADC). The JCMT maps have a median rms of 0.3\,K at a velocity resolution of 0.5\,\kms. Additionally, HCO$^+$(1--0) spectra observed with the IRAM 30m telescope as part of the Large Programme (LP017) for Master2 internships from Grenoble University, available at \url{https://www.iram.fr/ILPA/LP017/} were also used for comparison with the HCO$^+$(4--3) data.

\subsection{SEDIGISM}

We have used the data cubes for $J$=2--1 transition of \thCO\ and \CeiO\  at 220398.68 and 219560.36\,MHz respectively, that were observed as a part of the large-scale (84 deg$^2$) spectroscopic survey of the inner Galactic disk, named Structure, Excitation and Dynamics of the Inner Galactic Interstellar Medium \citep[SEDIGISM;][]{schuller2021}. These data were observed with the APEX telescope between 2013 and 2016 with an angular resolution of 30\arcsec\ and a 1$\sigma$ sensitivity less than 1.0\,K at
0.25\,\kms\ velocity resolution.  The SEGIDISM data were also used by \citet{Kuhn2022}.

\begin{figure*} 
\centering
\includegraphics[width=14.cm]{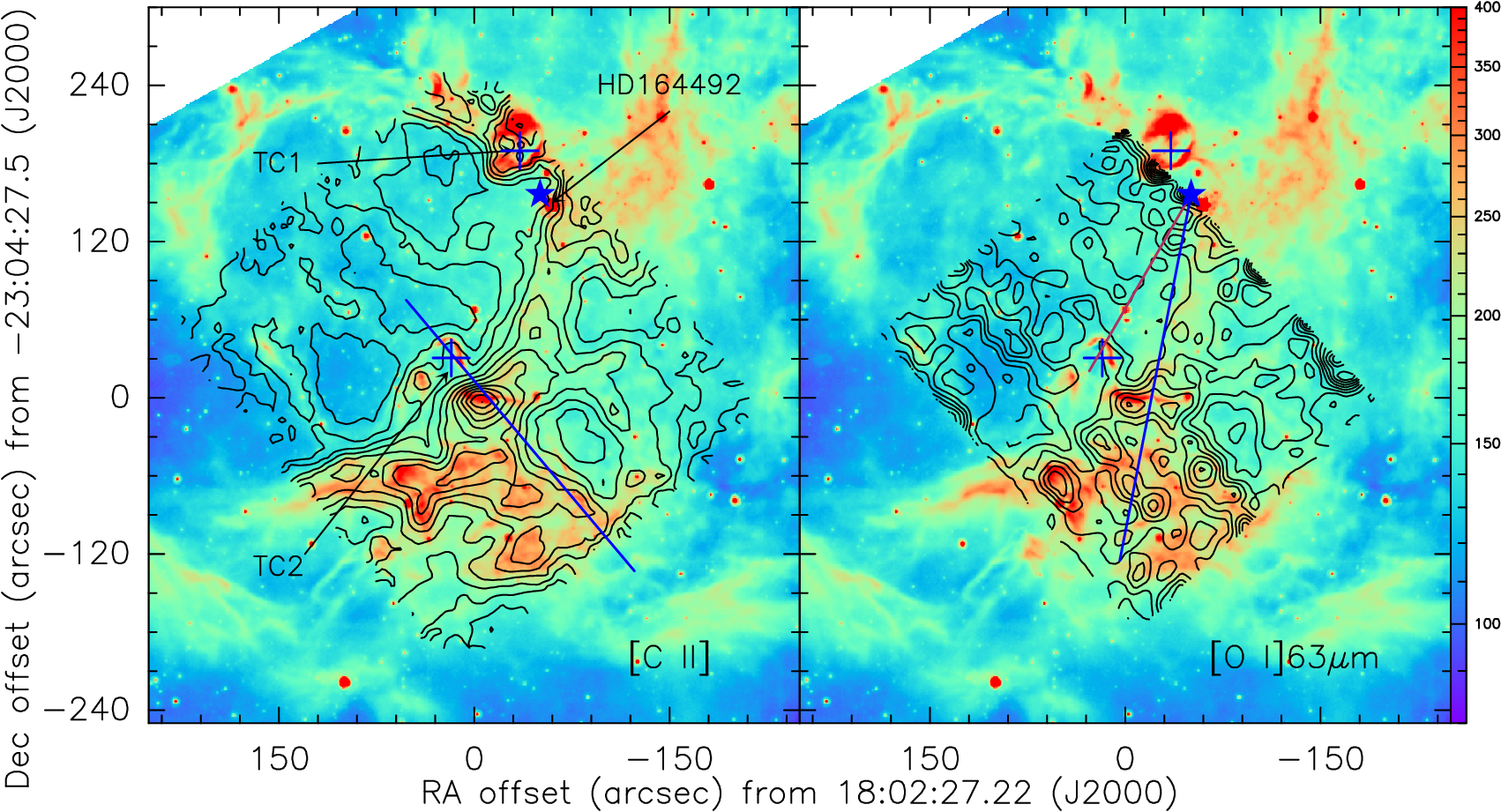}
\caption{Contours of \CII\ and \OI 63\,\micron\ intensities overlaid on color images of
8\,\micron\ Spitzer/IRAC region. Both maps are integrated between -4 to
14\,\kms. The \CII\ contours (in K\,\kms) correspond to 60 to 280 (in steps of 
20) and the contours for \OI 63\,\micron are at 0.5, 3 to 30 (in steps of 3), 35, 40, 45,
50. The \CII\ map is at the original resolution of 14\arcsec, and the \OI 63\,\micron\ map is smoothed to the same resolution. The colorscale  (in MJy\,sr$^{-1}$) for the 8\,\micron\ map is shown to the extreme right. The three solid straight lines correspond to directions along which position-velocity diagrams are analyzed (Fig.\,\ref{fig_cppv}). Other important sources in the region are  marked on
both panels and labeled on the the left panel. \label{fig_intmaps}}
\end{figure*}

\section{Molecular Gas Distribution in the region}

\begin{figure*}
\centering
\includegraphics[width=17.cm]{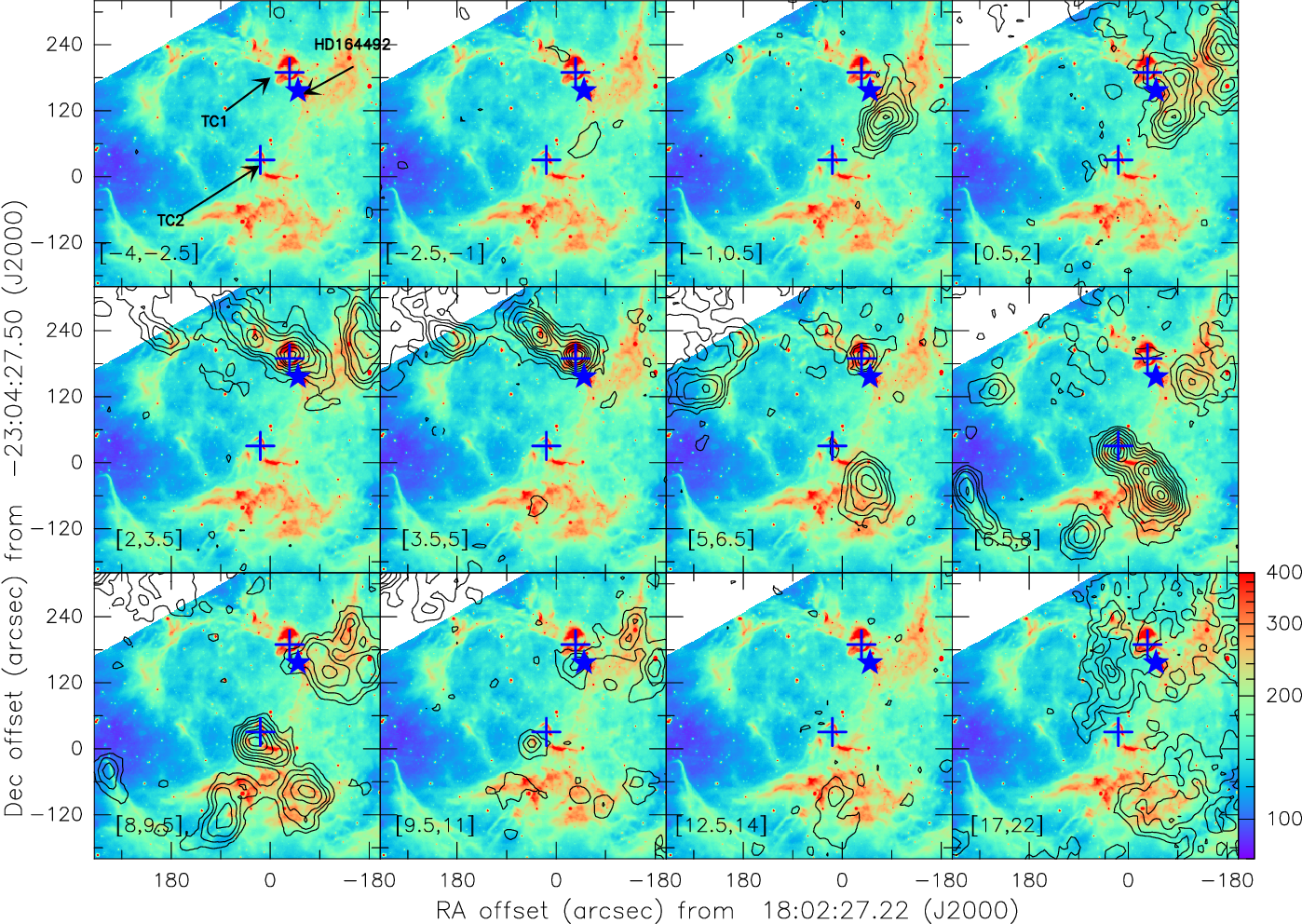}   
\caption{Map of \thCO(2--1) emission in 1.5\,\kms\ wide channels shown as contours overlaid on the Spitzer/IRAC 8\,\micron\ color image.  The numbers in square brackets in each panel indicate the velocity range over which the emission was integrated. The bottom rightmost panel is an exception that shows intensity integrated over a wider velocity range of 17 to 22\,\kms. The contours (in K\,\kms) are drawn at 1.5, 3, 4.5, 6, 7 to 28 in steps of 3. For the bottom right panel the levels (in K\,\kms) are at 6 to 18 in steps of  1. The color bar to the right shows the 8\,\micron\ intensity scale in units of MJy\,sr$^{-1}$.  HD164492 is marked with a star symbol and the sources TC1 and TC2 are marked with `+' signs.
\label{fig_13co21chanmap}}
\end{figure*}

Figures\,\ref{fig_13co21chanmap} and \ref{fig_co32chanmap} respectively, show the
velocity-channel maps for the \thCO(2--1) and \twCO(3--2) emission in this region. The \thCO(2--1) channel maps show emission mainly from the rims of the nebula,  from the pillar containing TC2 and at velocities $>$14\,\kms. The \twCO(3--2)
channel map  shows similar emission features except for the component
beyond 14\,\kms. The \thCO(2--1) SEDIGISM channel maps were also analyzed by
\citet{Kuhn2022}, who write that the \thCO\ emission is seen at almost any
velocity between -4 to 28\,\kms\ but the emission does not necessarily originate
from gas at the same distance. In the Galactic spiral arm model from Reid et al
(2016), gas near the Sagittarius Arm is expected to have \vlsr\ $\approx$
8\,\kms\ and gas in the near Scutum Arm would have \vlsr $\approx$ 18\,\kms. 
\citet{Kuhn2022} argued that given the 18\,\kms\ matches the expected velocity
of the near Scutum Arm it is likely that the CO emission in the velocity range
\vlsr $\approx$ 13--28\,\kms\ comes from a molecular cloud far behind Trifid.
These authors concluded that the multiple peaks seen in the average spectrum of the region lying between 0 to 13\,\kms\ are compatible with the acceleration of dense cloudlets during the shredding of the molecular cloud by the expanding \HII\ region created  by the radiative  feedback due to HD\,164492. They further noted that the observed velocities and structure are comparable to those obtained from numerical simulations of cloud disruption by radiative feedback \citep{Ali2018, Fukushima2020}. 

\section{Velocity Distribution of Diffuse and Dense PDR Gas}

The fine structure lines \CII\ at 158\,\micron\ and \OI\ at 63\,\micron\ are 
widely regarded as the bona-fide tracers of the PDR gas. Owing to the significantly different excitation conditions, the two spectral lines are also useful to discriminate between the diffuse and dense PDR gas. The \CII\ at 158\,\micron\ with an upper energy level of 91\,K and a critical density of 3000\,\cmcub\ is easily excited under average Galactic ISM conditions in which \cplus\ are found. The  \OI 63\,\micron\ line has an upper energy level of 227.7 K and a critical density of 2\,$\times 10^5$ \cmcub.  Thus, although atomic oxygen is as abundant as carbon the \OI 63 emission arises primarily from high density PDRs. Utilising the velocity-resolved spectra for both tracers we attempt to localize the diffuse and dense PDR gas in a velocity-coherent manner.

\begin{figure} 
\centering 
\includegraphics[width=9.cm]{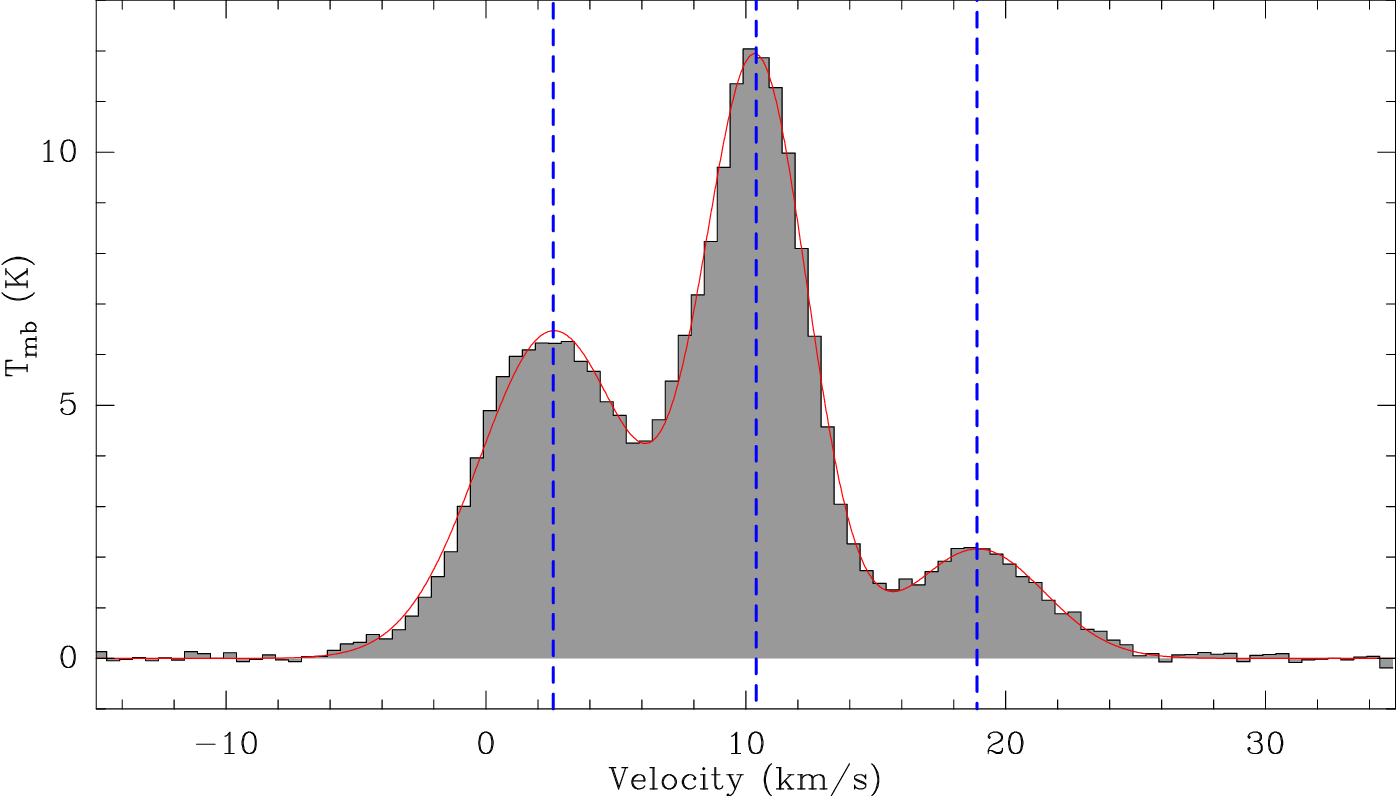}
\caption{The \CII\ spectrum averaged over the entire mapped region. The red
curve shows a fit to the spectrum using a three-component Gaussian profile
centered at 2.6, 10.4 and 18.9\,\kms. The centers of the Gaussian components are
marked by the vertical dashed lines. \label{fig_cpavg}} 
\end{figure}

Figure\,\ref{fig_intmaps} suggests an overall similarity between the
distribution of the velocity-integrated emission of \CII\ and \OI 63\,\micron\
when both are integrated between -4 to 14\,\kms. Both intensity maps trace the
southern part of the region hosting the pillar and the TC2 globule and part of
the the dust lanes seen in the optical image across M20. In order to investigate
the kinematics and distribution of the gas contributing to the emission of \CII\
we examine the \CII\ spectrum averaged over the entire mapped region
(Fig.\,\ref{fig_cpavg}). Using a multi-component Gaussian fit to the average
\CII\ spectrum we identify three distinct velocity components approximately
centered at 2.6, 10.4 and 18.9\,\kms\ -- these components  match well with the
velocity components already identified in the  molecular line data
(Fig.\,\ref{fig_cpavg}).

\begin{figure*} 
\centering
\includegraphics[width=16.cm]{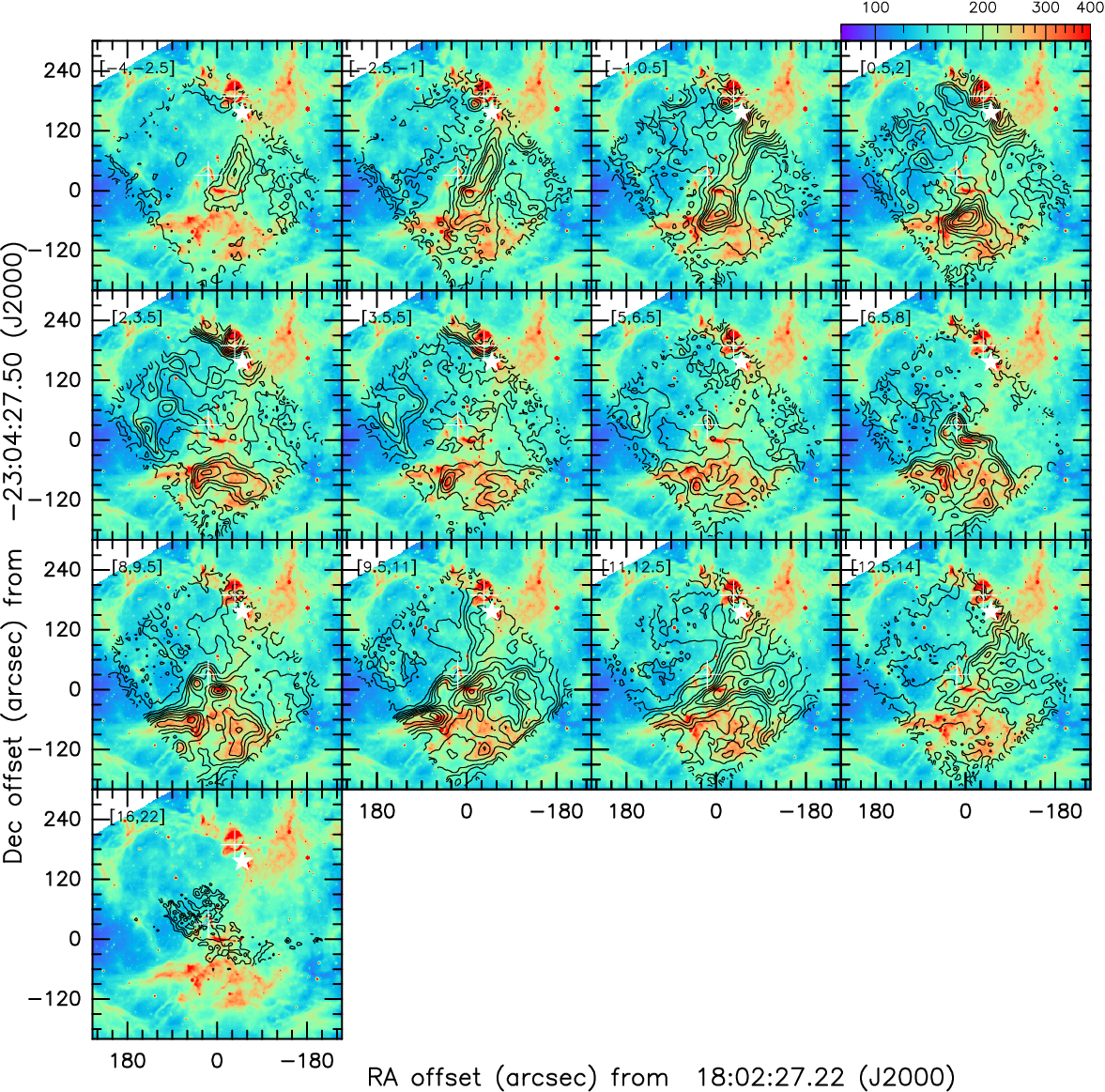}
\caption{Map of \CII\ emission in 1.5\,\kms\ wide channels shown as contours
overlaid on the Spitzer/IRAC 8\,\micron\ color image. The numbers in square brackets in each panel indicate the velocity range over which the emission was integrated. The contours are drawn at
0.5, 2.5, 4.5, 6 to 48 (K\,\kms) in steps of 6\,K\,\kms.  The lowermost panel is an exception that shows intensity integrated over a wider velocity range of 16 to 22\,\kms\ with contours drawn at 15 to 23\,K\,\kms\ in steps of 2\,K\,\kms. The color bar on the top shows the 8\,\micron\ intensity scale in units of MJy\,sr$^{-1}$. HD164492 is marked with a star symbol and the sources TC1 and TC2 are marked with `+' signs.
\label{fig_cpluschanmap}. } 
\end{figure*}

\begin{figure*} 
\centering
\includegraphics[width=16.cm]{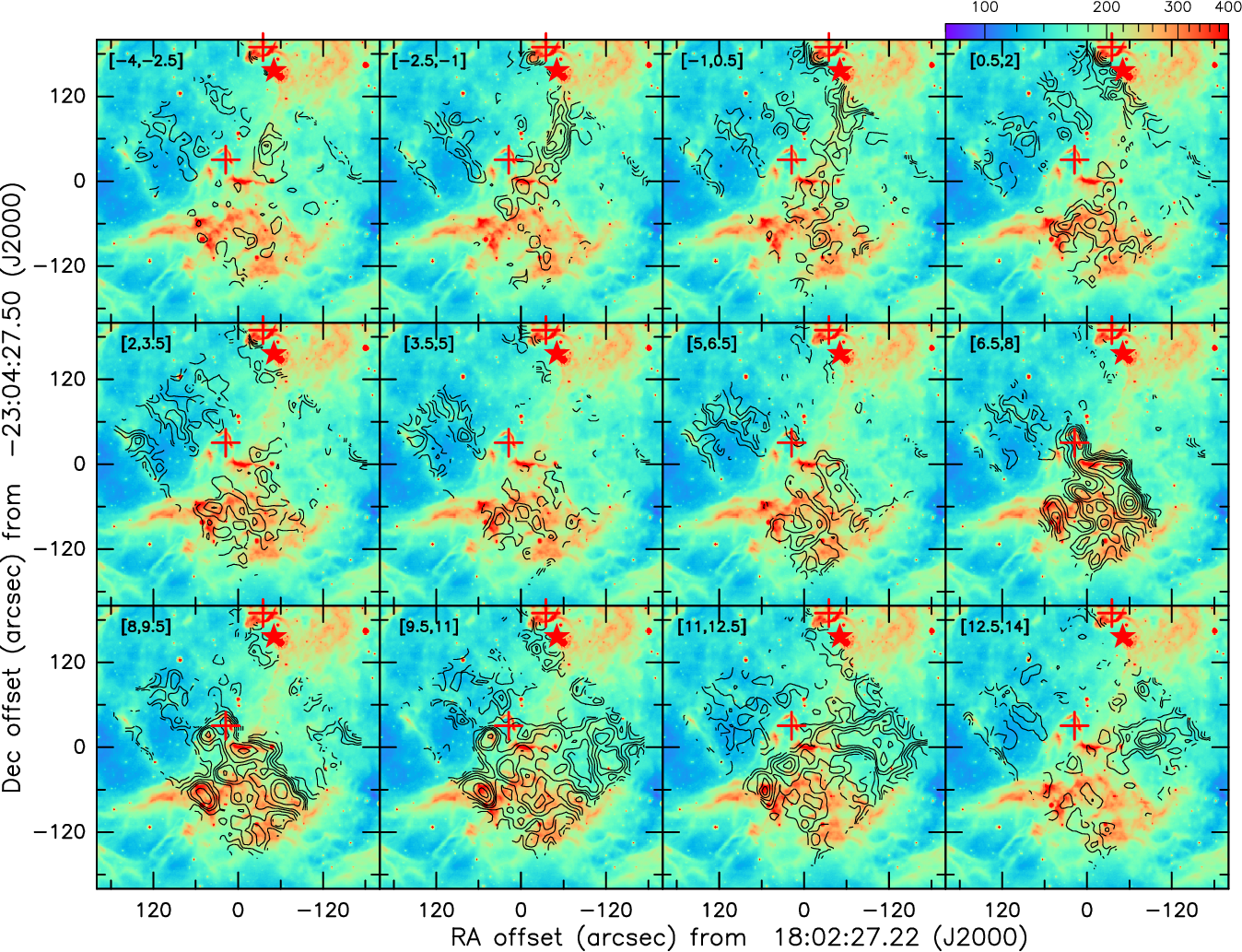}
\caption{Map of \OI 63\,\micron\ emission in 1.5\,\kms\ wide channels shown
both a color map and contours. The numbers in square brackets in each panel indicate the velocity range over which the emission was integrated.  The contours (in K\,\kms) are drawn at 0.5, 1.0,
1.75, 2.5, 3.25, 5 to 20 in steps of 1.  HD164492 is marked with an asterisk and
the sources TC1 and TC2 are shown as `+'.  The color bar on the top shows the 8\,\micron\ intensity scale in units of MJy\,sr$^{-1}$. HD164492 is marked with a star symbol and the sources TC1 and TC2 are marked with `+' signs.
\label{fig_o63chanmap}} 
\end{figure*}

\begin{figure*} \centering
\includegraphics[width=15.cm]{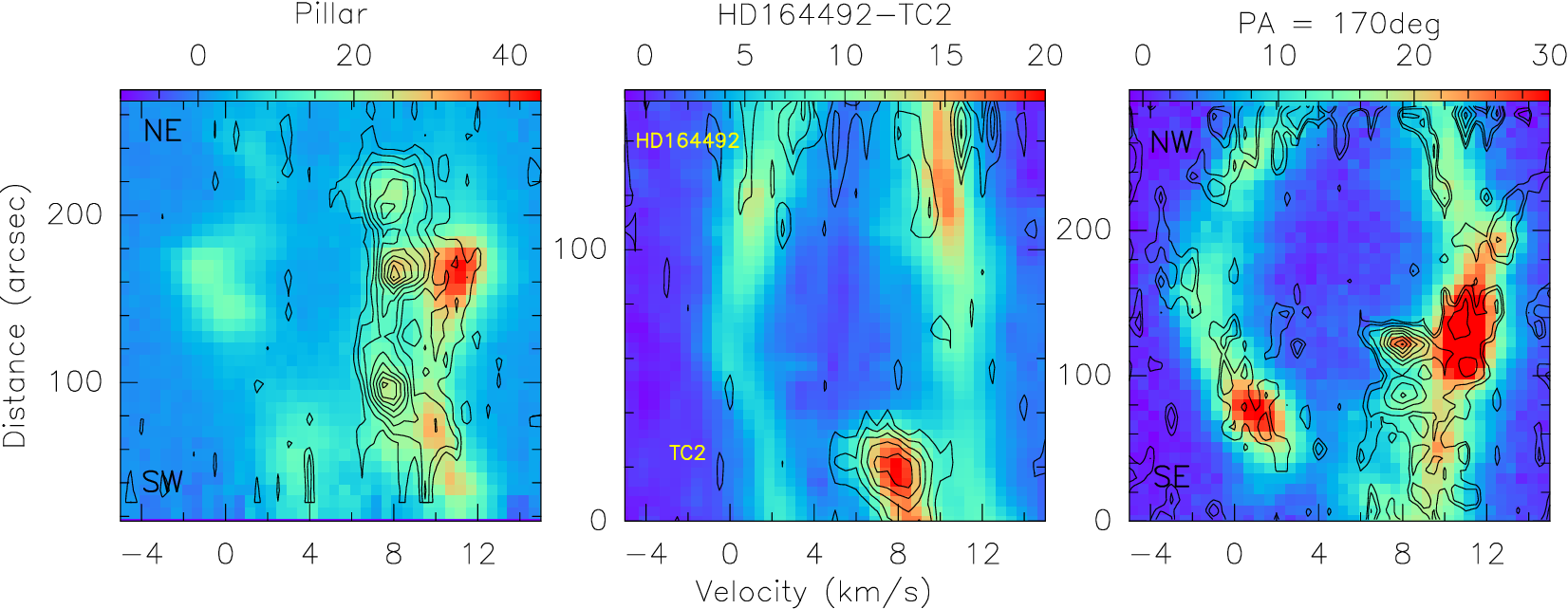}
\caption{Position-velocity diagrams of \CII\ emission along the cuts shown in
Fig.\,\ref{fig_intmaps}. The contours are for the $p$-v diagram for \OI 
63\,\micron\ and correspond to the levels (Left) 1 to 14\,K in steps of 1\,K, (Middle) 1 to 30\,K in steps of 1\,K, and (Right) 0.5, 1 to 35\,K in steps of 1\,K.  The colorbar on the top of each panel shows the \CII\ intensities in K. \label{fig_cppv}}
\end{figure*}

The distribution of the PDR gas is traced more clearly in the channel maps of
\CII\ and \OI 63\,\micron\ (Fig.\,\ref{fig_cpluschanmap}, \ref{fig_o63chanmap}). All the
three velocity components are seen in the \CII\ map. The \OI 63\,\micron\ map  primarily traces the emission of
the 8\,\kms\ component, but also shows some of the 2\,\kms\ velocity component,
in particular the PDR emission from the dark dust lane bisecting the image. 
This PDR emission stands out in both \CII\ and \OI 63\,\micron\ in the channels from -2.5 to
+0.5\,\kms, in contrast to the \thCO(2--1) and \twCO(3--2) emission which show very little
emission at velocities $<$ -1\,\kms. The velocity component at +18\,\kms\ is strongly contaminated by telluric \OI\ emission, which  masks any \OI\  emission in that velocity range.  The northern ridge containing TC1 was only partially observed in
the \OI 63\,\micron\ (Fig.\,\ref{fig_o63chanmap}).  The 8\,\kms\ component is clearly
visible in \CII, \OI 63\,\micron, \thCO(2--1) and \twCO(3--2) channel maps. Both  \CII\ and
\OI 63\,\micron\ emission trace the pillar-like structure, with the TC2 globule at its head,
at velocities between 5 to 9.5\,\kms, whereas the emission from the velocities
between 8 to 13\,\kms appear to be also tracing the denser gas located to the
south-west of the map. The velocity-resolved observations suggest that the \CII\
and \OI 63\,\micron\ emission as a function of velocity resemble emission from
different photo-irradiated layers of the molecular cloud impacted by the
expanding \HII\ region. Such a scenario is further substantiated by the
position-velocity ($p$-v hereafter) diagrams for both \CII\ and \OI 63\,\micron\
(Fig.\,\ref{fig_cppv}) along the three directions shown in the integrated
intensity  maps (Fig.\,\ref{fig_intmaps}). The clear circular shell generated in
the \CII\ $p$-v diagram by the 2\,\kms\ and 8\,\kms\ clouds is characteristic
of the position-velocity diagram due to shell created by an  expanding \HII\ region. 
Additionally we notice that for the direction along the pillar with TC2, while the
\OI 63\,\micron\ emission is centered at 8\,\kms, the \CII\ emission shows two branches
with the branch around 10\,\kms\ being brighter.  The \twCO(3--2)emission traces
warmer and denser gas than the \thCO(2--1) emission and though the features seen
in the position-velocity diagrams are closer to the structures seen in the \OI
63\,\micron, the blue-shifted half of the shell clearly detected in \CII\ is also partially detected in  \twCO(3--2) (Fig.\,\ref{fig_co32pv}).

\begin{figure*} \centering
\includegraphics[width=14.cm]{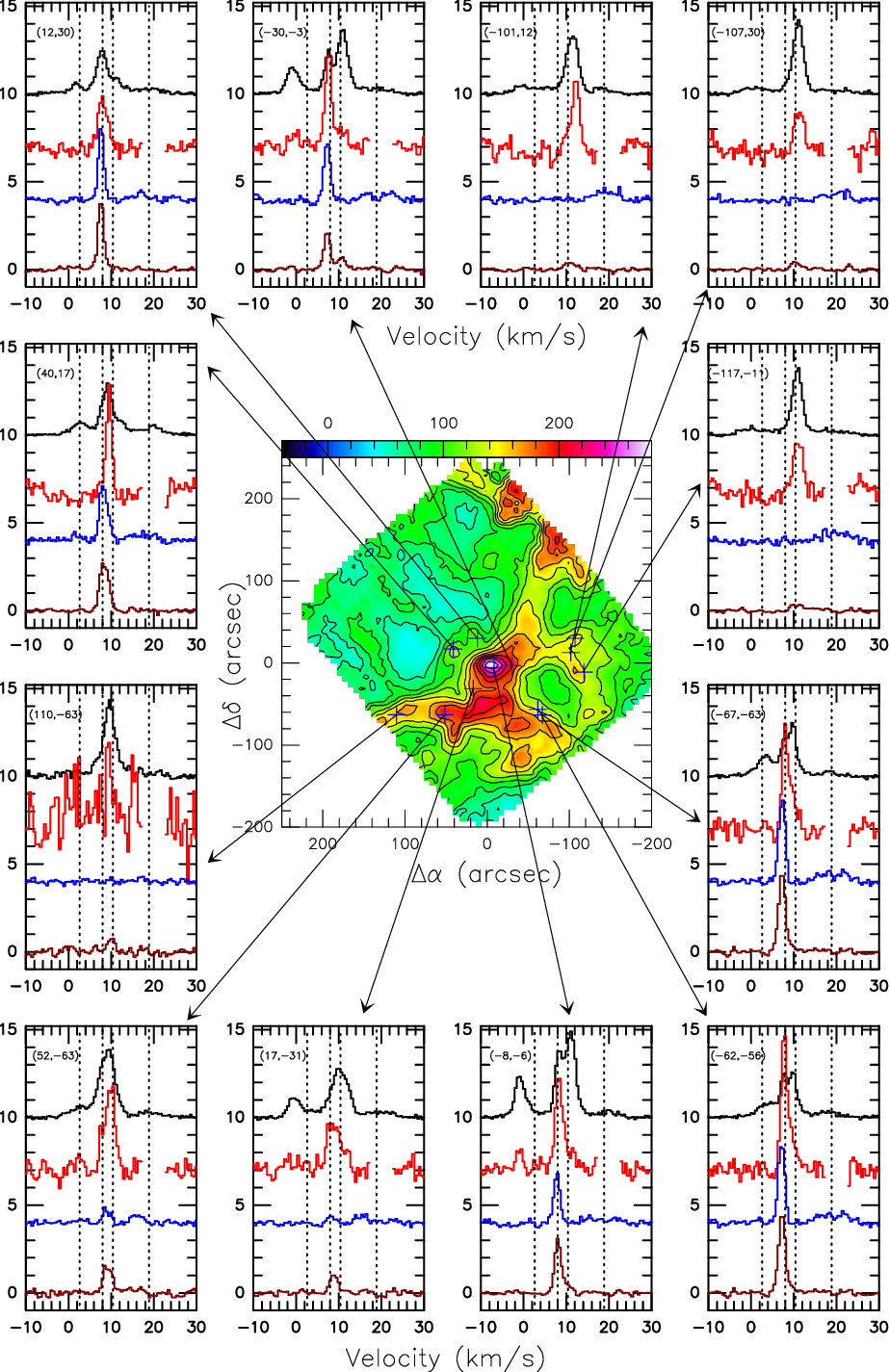}
\caption{Comparison of \CII\ (black), \OI 63\,\micron\ (red), \thCO(2--1) (blue) and
HCO$^+$(1--0) (maroon) spectra at selected positions. The positions are shown on
the \CII\ intensity map integrated over -3 to 14\,\kms. The contour levels in
intensity map correspond to 60 to 280 K\,\kms\ in steps of  20\,\kms\ and the
offsets in are with respect to the same center as in
Fig.\,\ref{fig_cpluschanmap}. The y-axis for each spectrum corresponds to
$T_{\rm mb}$. The \CII, \thCO(2--1) and HCO$+$(1--0) spectra are scaled by
factors of 0.11, 0.3 and 1.5 respectively. The \CII, \OI 63\,\micron\ and \thCO(2--1)
spectra are shifted along the y-axis for better visibility. Vertical dashed
lines correspond to 2, 8 and 18\,\kms, respectively. \label{fig_compspec}}
\end{figure*}

In order to explore whether the split in emission in the 8\,\kms\ cloud seen in
\CII\  close to the bright emission regions arises due to opacity
effects, we compared the \CII\ and \OI 63\,\micron\ spectra with \thCO(2--1) SEDIGISM
spectra and HCO$^+$(1--0) spectra at twelve selected positions (Fig.\,\ref{fig_compspec}). We find that at most positions where the
\CII\ spectra are double-peaked in the 8\,\kms\ component, the \OI 63\,\micron, 
\thCO(2--1)
and HCO$^+$ tend to align with one of the peaks and do not appear at the
velocity where the \CII\ spectra show the dip. Although restricted by the noise
level, the \CeiO(2--1) spectra at those positions also show the peak to be
aligned with one of the peaks, thus indicating that the two peaks correspond to
emission from clouds at two velocities. The overlapping but distinctly different
structures which manifest in the velocity ranges 6.5 to 8\,\kms\ and 9.5 to
11\,\kms\ in the \CII\ channel map strongly suggest that the two peaks in the
\CII\ spectrum arise from different layers of the PDR separated in velocity.
Fitting the 8\,\kms\ feature in the \CII\ spectra in this region using Gaussian
components we obtain velocities around 8.2 and 11\,\kms\ for the two components.
The \OI 63\,\micron\ spectra at these positions also show two components centered around
7.8 and 10.8\,\kms\ while the \thCO(2--1) profiles can be fitted with a single
component lying between 7 and 8.2\,\kms\ as  the RA offset ($\Delta \alpha$) changes from -42 to 28\arcsec. Of the two velocity components, the $\sim$ 7\,\kms\ component appears to be due to denser gas, since it is detected in both \OI 63\,\micron\ and HCO$^+$(1--0). The $\sim$ 11\,\kms\ component in contrast is dominantly seen in the \CII\ emission and hence is likely to be from more diffuse PDR gas.

\subsection{Gaussian Decomposition of Spectra}

We attempt to identify velocity-coherent features associated with the primary
velocity features in the \CII, \OI 63\,\micron\ and CO datasets. For this purpose we chose to use \CII\ because of its sensitivity to both diffuse and dense PDR gas, hence
tracing the overall kinematics better than the \OI 63\,\micron\ emission. We have used the
FUNStools.Decompose \footnote{\url{https://github.com/radioshiny/funstools}} to
decompose the \CII\ data cubes into multiple Gaussian components. The tool is
based on an algorithm that primarily decides the number of components and their
velocity positions in the smoothed spectrum using the first, second, and third
derivatives based on the  idea that a genuine velocity component is
continuous with the surroundings. The results of the first fitting are given as
the initial guess to a subsequent round of Gaussian fitting in an iterative
manner. The outcome are the parameters such as the central velocity and the
linewidth for each velocity component identified at each pixel in the \CII\
datacube. Subsequently, we identify coherent structures corresponding to each of
the velocity components so identified by applying the friends-of-friends (FoF)
technique to the central velocity of each of the components. The algorithm works
in an iterative manner in which it first selects the decomposed Gaussian seed
component with the maximum amplitude and gives the structure a number.
Subsequently the other Gaussian components in the neighboring pixels whose pixel
distance is less than $\sqrt{2}$ from the seed component are selected, and the
velocity differences of the seed and other components in the neighboring pixels
are checked. If the the velocity difference between the seed and other
components in the neighboring pixel is less than the velocity dispersion of the
seed ($\sigma_\nu^{\rm seed}$) the neighboring component is identified as a
friend of the seed and assigned the same group number.  If more than one
velocity component in a neighboring pixel is within the range of velocity
dispersion from the velocity of the seed, then only the closest one becomes the
friend of the seed. In the next iteration the assigned friends of the seed
become the seeds of the structure and the same procedure is repeated until there
are no more friends to assign.

\begin{figure} \centering
\includegraphics[width=0.45\textwidth]{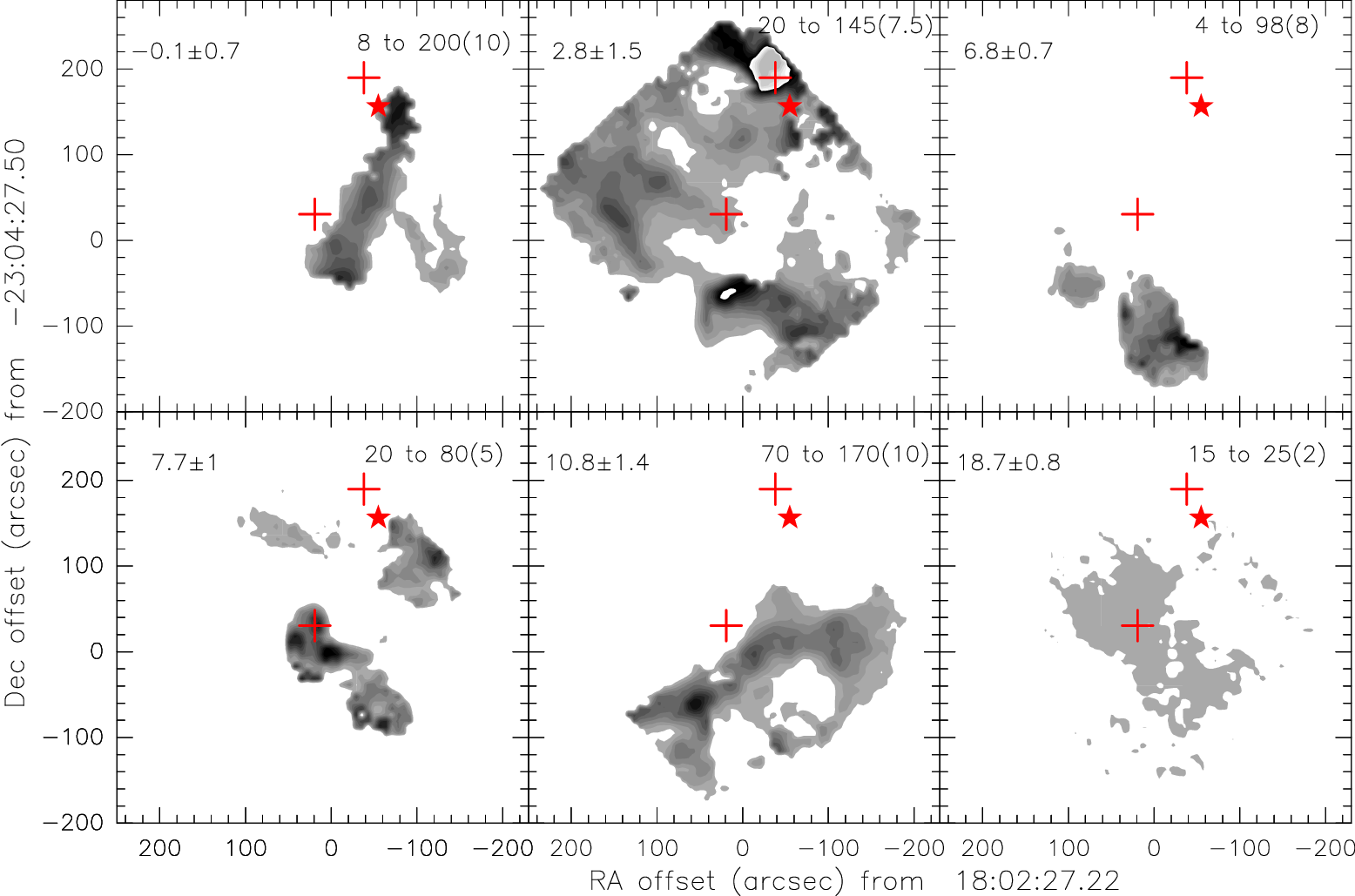}
\caption{Velocity coherent structures identified in the region based on the
\CII\ spectra using Gaussian decomposition followed by the identification of
groups using a friend-of-friend algorithm are shown as contours. The color image
corresponds to the velocity-integrated \CII\ emission and the colorscale is
shown at the top of the figure. In each panel we indicate the mean velocity of
the group along with the standard deviation in \kms\ on the left and the contour
levels in units of K\,\kms\ on the right. \label{fig_cpgrp}} \end{figure}

Using this method of identification of velocity-coherent structures in the \CII\
emission, we identify 5 main velocity components with velocities (in \kms)
centered at -0.1$\pm$0.7, 2.8$\pm$1.5, 6.8$\pm$0.7, 7.7$\pm$0.1, 10.8$\pm$1.4.
Additionally we detect the component at 18.7$\pm$0.8\,\kms, which as mentioned earlier
is unrelated to M20 and most probably associated with the Scutum arm. While the
outcome is consistent with the channel maps, decomposition of the spectra into
Gaussian components helps us to isolate the overlapping features. This is
particularly true for the features with mean velocities at 6.8, 7.7 and
10.8\,\kms.

A combination of the features seen in the channel maps of \CII, and \OI 63\,\micron\ and
grouping of the components identified in the Gaussian decomposition of \CII\
spectra provides us with a three-dimensional view of the multiple layers of PDR
shells associated with HD164492A.

\section{Discussion}

\subsection{Properties of the PDR shells}

The velocity-resolved spectra of \CII\ and \OI 63\,\micron\ have enabled us to identify
the red- and blue-shifted parts of the  far-ultraviolet (FUV; 6-13.6\,eV) irradiated molecular shell created by the expanding \HII\ region in M20. The structure seen particularly in the \CII\ $p$-v plots  (Fig.\,\ref{fig_cppv}) along selected directions suggests that the emission between -5 to 5\,\kms\ belongs to the blue-shifted (foreground) part of the shell, while the 5 to 15\,\kms\ emission corresponds to the red-shifted
part. The Gaussian decomposition of the spectra followed by identification of
velocity-coherent emission features identifies two and three sub-structures in
the blue- and red-shifted parts of the shell, respectively
(Fig.\,\ref{fig_cpgrp}). The detection of brighter \OI 63\,\micron\ emission from  the
red-shifted part of the shell indicates that it has higher density than the
blue-shifted part. Previous studies of Galactic PDRs have shown that the \CII\
emission is often optically thick, while the \OI 63\,\micron\ spectra are affected by
foreground low-excitation gas with atomic oxygen.  Since we did not detect
\thCII, which would have enabled us to the determine the optical depth of the
\CII\ emission, we instead use the results from \citet{Lefloch2002}, who
estimated that the kinetic temperature of the PDR gas was 300 K, to derive the
column density $N$(\cplus). Since it is likely that the 5--15 km/s PDR  emission
has high density, for this part of the shell we adopt an optical depth of 2, a value typically seen in Galactic PDRs \citep{Mookerjea2018}. We assume the -5 to +5 km/s PDR emission to  be optically thin. Equation (26) from \citet{Goldsmith2012} when modified for optically thick emission for the red-shifted shell, is as follows:

{\small
\begin{equation}
N(C^+) = {\rm 2.91\times 10^{15}\left[1+0.5{\rm
e^{91.25 K/T_{kin}}}\left(1+\frac{A_{ul}}{C_{ul}}\right)\right] \frac{\tau}{1-e^{-\tau}} \int T_{\rm mb}d\upsilon}
\end{equation}
}
where $A_{\rm ul}$ = 2.3$\times 10^{-6}$\,s$^{-1}$, \tkin\ is the gas kinetic
temperature, the collision rate is $C_{\rm ul}$ = $R_{\rm ul}n$ with $R_{\rm
ul}$ being the collision rate coefficient with H$_2$ or H$^0$, which depends on
$T_{\rm kin}$, and $n$ is the volume density of H. Since the critical density of
the \CII\ transition is  $n_{\rm cr}$=3000\,\cmcub, and since it is likely that
most of  the  \CII\ detected could be at such densities along with some emission
arising from clumps with densities exceeding 10$^5$\,\cmcub, we assume a density
of  10$^4$\,\cmcub\ to estimate the $N$(\cplus) density distribution of the
PDR gas. For this calculation we consider excitation of \CII\ due to 
\cplus--H$_2$ collisions, with  $R_{ul}$ = 3.8$\times
10^{-10}$\,cm$^3$\,s$^{-1}$. The value of $N$(\cplus) estimated for the
red-shifted shell is between (1--3)$\times 10^{18}$\,\cmsq, while for the
blue-shifted shell $N$(\cplus) ranges between (2--9)$\times 10^{17}$\cmsq. Based
on LWS observations with a beam size of 45\arcsec\ \citet{Lefloch2002} had
estimated that $N$(\cplus) in the PDR of the globule is around
(4.5--5.7)$\times 10^{17}$\,\cmsq, after subtracting a 30\% contribution to \CII\
emission from the ionized gas in the \HII\ region.  At TC2, we estimate a total
$N$(\cplus) of 1.2$\times 10^{18}$\,\cmsq\ and if 30\% of the \CII\ emission
were to arise from the ionized gas, the contribution of the PDR  will be $\sim
8.4\times 10^{17}$\,\cmsq. Within the limits of the uncertainties  this estimate
is consistent with the values determined by \citet{Lefloch2002}.

In order to examine the properties of the PDRs created in the FUV irradiated
parts of the red- and blue-shifted molecular shells we derive the \CII/\OI
63\,\micron\ intensity ratios for both components. The area over which the ratio can be
determined is restricted by the usable pixels of the \OI 63\,\micron\ map. For the red and
the blue-shells the \CII/\OI 63\,\micron\ intensity ratios (in energy units) vary
between (0.8 to 2.5) and (1 to 1.5), respectively. \citet{Lefloch2002} estimated
the FUV field intensity at the location of the globule TC2 due to HD 164492A to
be $G_0=1000$. Comparison of the observed \CII/\OI 63\,\micron\ intensity ratios with the
results of the Kaufman PDR model \citep{Kaufman2006} for an FUV intensity of
$G_0$=1000 suggest densities of 4$\times 10^3$ and 500\,\cmcub\ for the red and
the blue-shifted shells. Based on the detection of \OI 63\,\micron\ emission, it appears
that these densities are lower limits. The lower derived values of the density
could be because the volume and hence beam-filling factor of the \CII\ emission
is likely to be significantly larger than the \OI 63\,\micron\ emission that arise from high
density PDRs and this could lead to higher values of observed \CII/\OI 63\,\micron\
intensity ratios, which imply lower densities. At the position corresponding to
TC2 (40\arcsec, 17\arcsec) we have fitted the \OI 63\,\micron\ spectrum using a single
velocity component Gaussian profile resulting in $v_{\rm LSR}$ = 9.5\,\kms\ and
$\Delta v$ = 1.6\,\kms. The \CII\ spectrum at this position between -8 to
17\,\kms\ shows three velocity components, hence we have fitted it with three
Gaussians. While fitting the \CII\ profile we kept the center and the width of one of the components fixed at the fitted values obtained for the \OI 63\,\micron\ profile. Thus, for this velocity component, which is seen in both the \CII\ and \OI 63\,\micron\ emission, we estimate a \CII/\OI 63\,\micron\ intensity ratio (in energy units) of 0.2. For such a ratio and a G$_0$ = 1000, the PDR models indicate a density of 10$^4$\,\cmcub.

\subsection{Kinematics of the PDR shells}

Based on the \CII\ intensities, we estimate the total integrated column
densities ($N$(\cplus)) of the blue- and the red-shifted shells to be 5.9$\times
10^{20}$ and 2.4$\times 10^{21}$\,\cmsq, respectively. Considering a C/H$_2$
ratio of 1.5$\times 10^{-4}$ at the distance of M20 (1180\,pc), the masses of
the blue- and red-shifted shells are estimated to be 100 and 416\,\msun,
respectively. Based on the shells identified in the position-velocity plots of
\CII\ (Fig.\,\ref{fig_cppv}), we estimate the median expansion velocity of the
shells to be $\sim 5$\,\kms. Thus the total kinetic energy of the PDR shells is 
approximately 1.3$\times 10^{47}$ erg. \citet{ChaissonWilson1975} derived a few
parameters for the \HII\ region M20, created by the star HD169942A: it extends
over 5\farcm3$\times$6\farcm4, which at a distance of 1180\,pc corresponds to a
size of 1.8$\times$2.2\,pc, has an average electron density ($n_{\rm e}$) of
100-150\,\cmcub\ and has an electron temperature ($T_{\rm e}$) of 8150\,K.
\citet{GarciaRojas2006} estimated the electron number density to be
270$\pm60$\,\cmcub. The total thermal energy of the \HII\ region written as
${\rm E_{th} = \frac{3}{2} n_e \frac{4\pi}{3}\,R^3 k_B T_{\rm e}}$, where ${R}$
is  the radius of the \HII\ region and $k_{\rm B}$ is the Boltzmann constant.
Based on these parameters we estimate the total thermal energy of the \HII\
region to be 6$\times 10^{46}$\,erg. Thus, subject to uncertainties of the
electron density in the \HII\ region, the kinetic energy of the PDR shell is
about two times the thermal energy of the \HII\ region. 

The photoionized gas in the interior of an expanding \HII\ region exerts a
pressure force and delivers outward radial momentum and kinetic energy to the
swept-up shell. For a spherical \HII\ region the momentum delivered to the
ambient medium is \citep{Krumholz2017}:

\begin{equation} p_{\rm sh} = 1.5\times 10^5 \left[\frac{n_{\rm
H}}{10^2}\right]^{-1/7} \left[\frac{T_e}{10^4}\right]^{-8/7} \left[\frac{N_{\rm
Lyc}}{10^{49}}\right]^{4/7} \left[\frac{t}{10^6}\right]^{9/7} \, M_\odot\mbox{
km s}^{-1}. \end{equation}

where n$_{\rm H}$ is the number density of hydrogen in the ambient medium,
$T_{\rm e}$ is the electron temperature, $N_{\rm Lyc}$ is the Lyman continuum
photon rate due to HD164492A and $t$ is the age of the \HII\ region.
\citet{Lefloch2000} estimated the density of the shell to be 10$^5$\,\cmcub\ and
mean density of the cloud to be 2$\times 10^{3}$\,\cmcub. We note that the
assumption of the \HII\ region expanding in a uniform density environment could
lead to uncertainties in the determination of its age and subsequently in the
value of the density as well. In order to include the effects of higher density
material that the expanding \HII\ region would necessarily encounter in an
otherwise clumpy molecular cloud, we assume the ambient density to be
10$^4$\,\cmcub.  For HD 164492A, an O7V type star the Lyman continuum photon
rate is $5.6\times 10^{48}$\,s$^{-1}$ \citep{Martins2005}. Using Eq. (2) for the
estimated age of the M20 nebula of  0.5\,Myr we estimate the total momentum of
the PDR shell to be 2.8$\times 10^4$\,\msun\ \kms. From the observations we
estimate the momentum of the PDR shell with a mass of 516\,\msun\ moving with a
velocity of 5\,\kms\ to be $\sim 2.6\times 10^3$\,\msun\,\kms,
which is lower by a factor of 10 than the $p_{\rm sh}$ estimated from Eq.(2). This
could be due to the molecular (and photodissociated) material  which might have
been dispersed, hence carrying away part of the momentum.

The kinetic energy of the swept up shell is given by \citep{Krumholz2017}:

\begin{equation}
 E_{\rm sh} =  8.1\times 10^{47} \left[\frac{n_{\rm H}}{10^2}\right]^{-10/7} \left[\frac{T_e}{10^4}\right]^{10/7} \left[\frac{N_{\rm Lyc}}{10^{49}}\right]^{5/7} \left[\frac{t}{10^6}\right]^{6/7} \,\mbox{erg}  
\end{equation}

The kinetic energy of the swept up shell estimated using Eq. (3) is 3.1$\times
10^{44}$ erg, which is much smaller than the observed value by a factor of more
than 320. It is clear that a simplistic model of an \HII\ region expanding
thermally in a uniform density environment cannot provide the observed kinetic
energy of the shell. on the other hand, the wind luminosity from HD169942A 
is estimated to be 1.7$\times 10^{36}$\,erg\,s$^{-1}$ \citep{Rho2008, Howarth1989,Prinja1990}, indicating that the wind  has pumped in a total kinetic energy of 5.2$\times 10^{46}$\,erg over the 0.5\,Myr lifetime of the Trifid nebula. This number matches quite well with the observed kinetic energy of the PDR shells as well as with the thermal energy of the ionized gas in the nebula.

%A possible explanation of the deficit in measured kinetic energy is that most of %the energy due to the expansion of the \HII\ region got dumped in the molecular %cloud beyond the PDR shell. 

\section{Summary \& Conclusions}
%{\bf NOT UPDATED YET}
We have analysed the distribution of the PDR surfaces created by the FUV radiation from the O-type star HD\,164492 that illuminates the \HII\ region associated with the M20 nebula based on decomposition of the \CII\ emission into velocity components. We disentangle the high density PDR material using the \OI 63\,\micron\ emission.   The structure of the PDRs is consistent with shells created by the  \HII\ region  expanding approximately with a velocity of 5\,\kms. Based on the \CII/\OI 63\,\micron\ intensity ratios we estimate the densities of the red and the blue-shifted parts to be 4000 and 500\,\cmcub, which are likely to be the lower limits considering the possible mismatch between the beam-filling factors of the \CII\ and the \OI 63\,\micron\ emission. The ratio, when estimated by considering the intensities of velocity components seen in both the \CII\ and \OI 63\,\micron\ emission, suggest a density of 10$^4$\,\cmcub, which is consistent with the ambient PDR density determined by \citet{Lefloch2002}. The total kinetic energy estimated for the shells, $\sim 10^{47}$\,erg, is consistent with both the thermal energy of the \HII\ region as well as the total kinetic energy pumped in by the stellar wind during the lifetime (0.5\,Myr) of the nebula. This is also an indirect confirmation of the age of the nebula. The total momenta of the detected parts of the shell is about 2.6$\times 10^3$\,\msun\,\kms, which is about ten times less than the momentum that a spherical \HII\ region with the physical parameters of M20 can deliver. The presence of ``holes" on either side of the dark lanes, which essentially are remnants of the blue-shifted part of the shell, also suggest that a significant part of this shell (molecular and/or PDR) has been blown away by the \HII\ region and by the stellar wind. The material thus lost would have carried with it significant momentum and hence explains the discrepancy between the measured momentum of the shell and the momentum that the \HII\ region is likely to deliver.

Radiative and mechanical feedback due to the massive stars both at the end of their lives  (in the form of supernovae) and in their youth (as expanding \HII\ regions and stellar winds) influence the triggering and/or hindrance to the formation of new generation of stars.  The relative importance of the \HII\ regions and the stellar winds in shaping the structure of the ambient interstellar medium is of interest in the context of the evolution of the Galactic interstellar medium and star formation scenario \citep[cf.][and references therein]{Pabst2020,Tiwari2021}. This work presented observational evidence of the role of the stellar winds in shaping the structure of the putative sites of triggered star formation in the Trifid nebula. The velocity information in the \CII\ data enabled us to quantitatively compare the stellar wind inputs with the kinetic energy of the PDR shells and confirm that in this case the stellar winds provide the energy for expansion rather than the thermal expansion of the \HII\ region. For a more comprehensive understanding of the kinematics of the M20 nebula, extending \CII\ map to the north and obtaining a larger \OI 63\,\micron\ map  would enable a quantitative analysis of TC1 similar to the analysis done by \citet{Lefloch2002} for TC2. Such studies are crucial to obtain an improved understanding of the role of stellar feedback in determining the structure of the ambient medium that possibly trigger the formation of a new generation of stars in the region. 

\begin{acknowledgments}
%\acknowledgements
BM acknowledges the support of the Department of Atomic Energy, Government of India, under Project Identification No. RTI 4002 and the software related discussions with Akhil Lasrado. This research has made use of the NASA/IPAC Infrared Science Archive, which is funded by the National Aeronautics and Space Administration and operated by the California Institute of Technology. This publication is based on data acquired with the Atacama Pathfinder Experiment (APEX) under programmes 092.F-9315 and 193.C-0584. APEX is a collaboration among the Max-Planck-Institut fur Radioastronomie, the European Southern Observatory, and the Onsala Space Observatory. The processed data products are available from the SEDIGISM survey database located at \url{https://sedigism.mpifr-bonn.mpg.de/index.html}, which was constructed by James Urquhart and hosted by the Max Planck Institute for Radio Astronomy. This research used the facilities of the Canadian Astronomy Data Centre operated by the National Research Council of Canada with the support of the Canadian Space Agency. This work is based [in part] on observations made with the Spitzer Space Telescope, which is operated by the Jet Propulsion Laboratory, California Institute of Technology under a contract with NASA. This dataset or service is made available by the Infrared Science Archive (IRSA) at IPAC, which is operated by the California Institute of Technology under contract with the National Aeronautics and Space Administration.
\end{acknowledgments}

\newpage
\appendix{}
\restartappendixnumbering
%\appendix

Here we present additional CO(3-2) channel maps and position-velocity diagrams.

\section{Channel map and Position-Velocity Plots for \twCO(3--2)}
\begin{figure*}[h]
\centering
\includegraphics[width=17.cm]{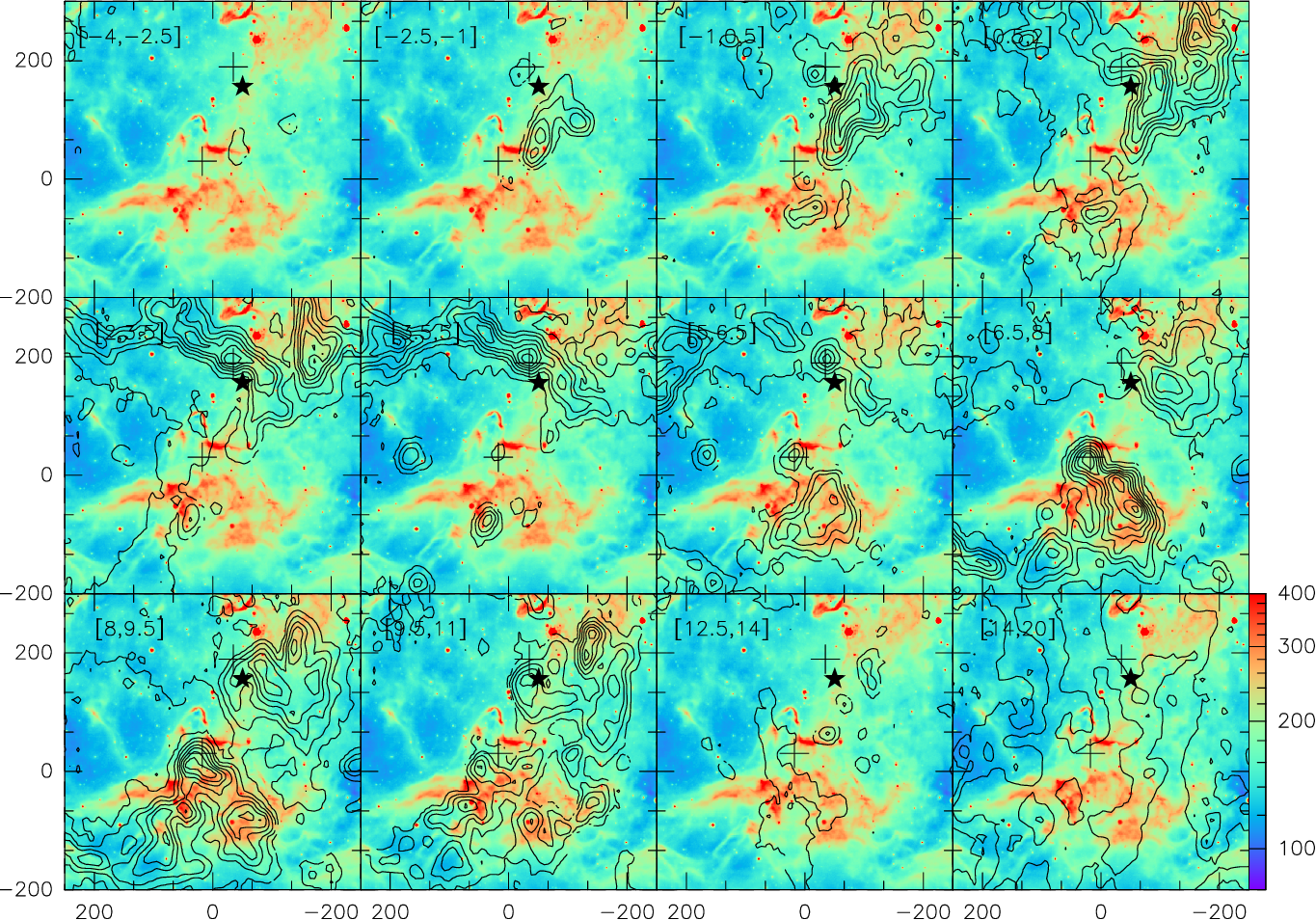}
\caption{Map of \twCO(3--2) emission in 1.5\,\kms\ wide channels shown as contours overlaid on the Spitzer/IRAC 8\,\micron\ color image.  The numbers in square brackets in each panel indicate the velocity range over which the emission was integrated. The bottom  rightmost panel is an exception that shows intensity integrated over a wider velocity range of 14 to 20\,\kms. The contours are drawn at 0.5, 2.5, 4.5, 6 to 48 (K\,\kms) in steps of 6\,K\,\kms. The color bar to the right shows  the 8\,\micron\ intensity scale in units of MJy\,sr$^{-1}$. HD164492 is marked with a star symbol and the sources TC1 and TC2 are marked with `+' signs.
\label{fig_co32chanmap}}
\end{figure*}
\begin{figure*}
\centering
\includegraphics[width=15.cm]{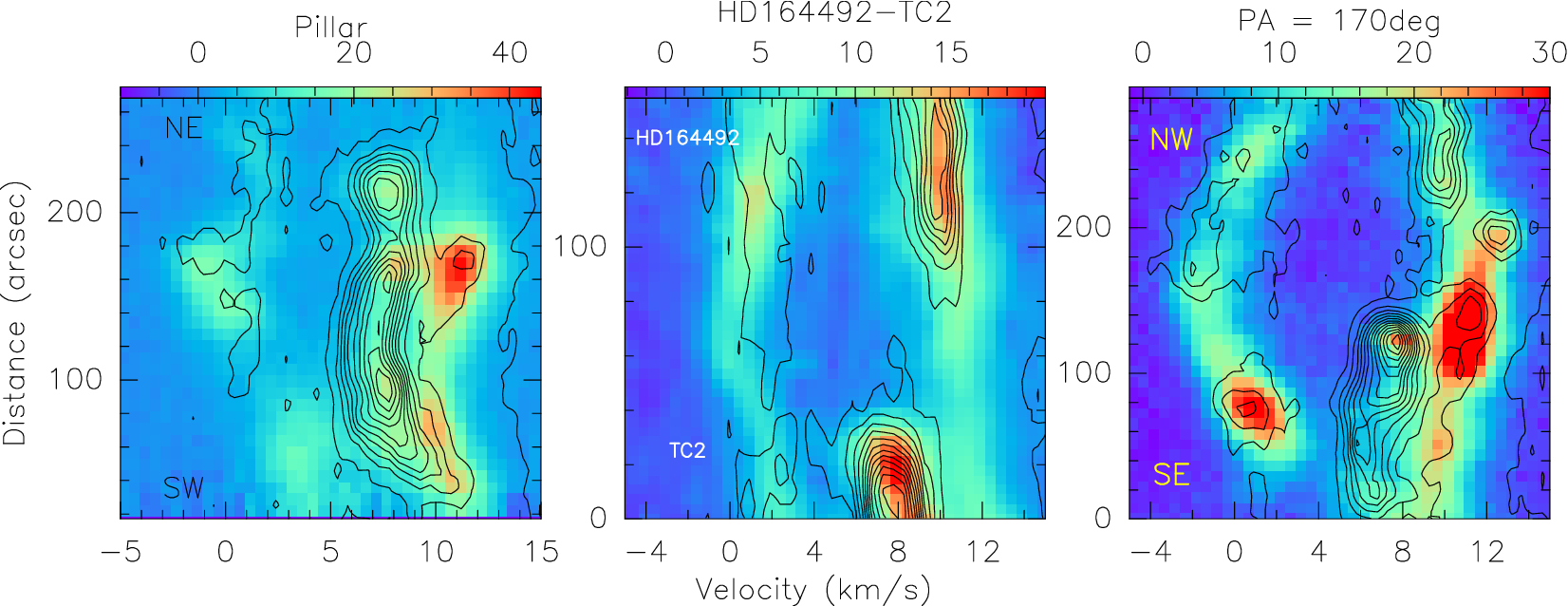}
\caption{Position-velocity diagrams of  \twCO(3--2) emission along the cuts shown in Fig.\,\ref{fig_cpluschanmap} shown as contours overplotted on the color image of the \CII\ $p$-v diagram.  The contour levels are for $T_{\rm mb}$ from 1 to 32\,K in steps of 3\,K  for the left panel and from 1 to 27\,K in steps of 2\,K for right panel. The colorbar on the top of each panel shows the \CII\ intensities in K.
\label{fig_co32pv}}
\end{figure*}

\facility {IRSA, Spitzer, Herschel, SOFIA}

\bibliographystyle{aasjournal}{}
\bibliography{m20}

%\label{lastpage}
\end{document}